\documentclass[11pt]{article}
\usepackage{moriond,epsfig}
\usepackage{wrapfig}
\usepackage{subfigure}

\def\beq{\begin{equation}}
\def\eeq{\end{equation}}
\def\bea{\begin{eqnarray}}
\def\eea{\end{eqnarray}}

\def \prd#1#2#3{{Phys. Rev. D} {\bf#1}, #2 (#3)}
\def \prl#1#2#3{{Phys. Rev. Lett.} {\bf#1}, #2 (#3)}
\def \plb#1#2#3{{Phys. Lett. B} {\bf #1}, #2 (#3)}

\def \npb#1#2#3{{Nucl. Phys. B} {\bf #1} #2 (#3)}

\def \etal{{\it et\,al.}}

\def \asy#1#2{{^{+#1}_{-#2}}}
\newcommand{\zero}{\mbox{$^{0}$}}			
\newcommand{\plus}{\mbox{$^{+}$}}
\newcommand{\minus}{\mbox{$^{-}$}}

\newcommand{\goesto}{\mbox{$\rightarrow$}}

\newcommand{\bbar}{\mbox{$\bar{b}$} }
\newcommand{\bbbar}{\mbox{$b\bbar$}}

\newcommand{\psip}{\mbox{$\psi(2S)$}}
\newcommand{\jpsi}{\mbox{J/$\psi$}}

\newcommand{\upsi}{\mbox{$\Upsilon(1S)$}}
\newcommand{\upsii}{\mbox{$\Upsilon(2S)$}}
\newcommand{\upsiii}{\mbox{$\Upsilon(3S)$}}

\newcommand{\upsns}{\mbox{$\Upsilon(nS)$}}
\newcommand{\upsms}{\mbox{$\Upsilon(mS)$}}

\newcommand{\chib}{\mbox{$\chi_{b}(1P)$}}
\newcommand{\chibp}{\mbox{$\chi_{b}(2P)$}}

\newcommand{\piz}{\mbox{$\pi$}^{0}}

\newcommand{\pipi}{\mbox{$\pi$}^{+}\mbox{$\pi$}^{-}}
\newcommand{\pizpiz}{\mbox{$\pi$}^{0}\mbox{$\pi$}^{0}}

\newcommand{\ee}{\mbox{$e\plus e\minus$}}
\newcommand{\gamgam}{\mbox{$\gamma\gamma$}}
\newcommand{\mumu}{\mbox{$\mu\plus\mu\minus$}}

\begin{document}
\vspace*{4cm}
\title{BOTTOMONIUM AND CHARMONIUM RESULTS FROM CLEO}

\author{ T. K. PEDLAR }

\address{Department of Physics, Luther College, 700 College Drive\\
Decorah, IA, 52101, USA}

\maketitle\abstracts{Heavy Quarkonium Physics continues to be a focus of the
work done by the CLEO Collaboration.  We present several results in the spectroscopy
of both bottomonium and charmonium systems using CLEO's data sets taken at the
$\upsiii$, $\upsii$ and $\psip$ resonances. 
}

\section{Introduction}  In 2000, CLEO stopped running at the $\Upsilon(4S)$ 
for B-meson studies, and began an eight-year study of the states of 
bottomonium, charmonium and open charm mesons, the latter two of which 
were performed primarily as the experiment evolved into CLEO-c.  We 
present here several results from our full charmonium and
bottomonium data sets.

\section{Hadronic Transitions}
The study of hadronic transitions among heavy quarkonium states provides important tests
for non-perturbative Quantum Chromodynamics (QCD)~\cite{kqrr}.   In the
multipole expansion,~\cite{gottfried} 
hadronic transitions among heavy quarkonium states proceed by emission and
hadronization of soft gluons.   
The non-relativistic nature of the bottomonium 
system and the richness of the spectrum of bound states make it an excellent laboratory for the study of the low-$q^2$ hadronization process.  

\subsection{$\pi\pi$ Transitions}

First, we report ~\cite{dipibfpaper} improved measurements of the branching fractions for $\pi\pi$ transitions among the vector
states of the bottomonium system.
Dipion transitions from $\upsiii$ to the lower vector states ($\upsii,\;\upsi$) and from $\upsii$ to $\upsi$ 
have been of interest ever since their first observation in 
1982.  There has recently been a resurgence of interest in dipion transitions following the 
observation of new $\pipi$ transitions
by several experiments.  
Additional motivation to update measurements of 
the branching fractions for bottomonium dipion transitions 
comes from the prospects of using $\upsiii,\upsii\goesto\pi\pi\upsi$ as a clean
source of tagged $\upsi$ to study exclusive $\upsi$ decays, including searches for invisible decay modes.

 \begin{wraptable}[9]{r}{4in}
\caption{Results of improved branching fraction measurements for the
processes $\upsns\goesto\pi\pi\upsms$.~\label{dipitable}
}
\small
\begin{tabular}{|l|l|l|}
\hline
Mode & BF (\%) & PDG BF (\%)\\
\hline 
$\upsiii\goesto\pipi\upsi$ & $4.46\pm 0.01\pm 0.13$ & $4.48\pm 0.21$\\
$\upsiii\goesto\pizpiz\upsi$ & $2.24\pm 0.09\pm 0.11 $ & $2.06\pm 0.28$\\
$\upsiii\goesto\pizpiz\upsii$ & $1.82\pm 0.09\pm 0.12 $ & $2.00\pm 0.32$ \\
\hline
$\upsii\goesto\pipi\upsi$  & $18.02\pm 0.02\pm 0.61 $ & $18.8\pm 0.6$ \\
$\upsii\goesto\pizpiz\upsi$ & $8.43\pm 0.16\pm 0.42$ & $9.0\pm 0.8$\\
\hline
\end{tabular}
\end{wraptable}

In this analysis, we study the transitions both inclusively (in which case we detect
only the pair of charged pions) and exclusively (in which case we detect, in addition
to the charged or neutral pair of pions, the decay of the daughter $\upsns$ state to either
$\mumu$ or $\ee$).  In each case, the 
primary quantity used to identify our observation of the dipion transitions of
interest is mass recoiling against the dipion system.  From the recoil
mass histograms, yields for each process may be obtained and converted to the branching
fractions presented in Table~\ref{dipitable}.  In every case the branching fractions obtained
are more precise than the current PDG~\cite{pdg} world average.

\begin{wrapfigure}[13]{r}{0.55\linewidth}
\includegraphics[width=0.45\linewidth]{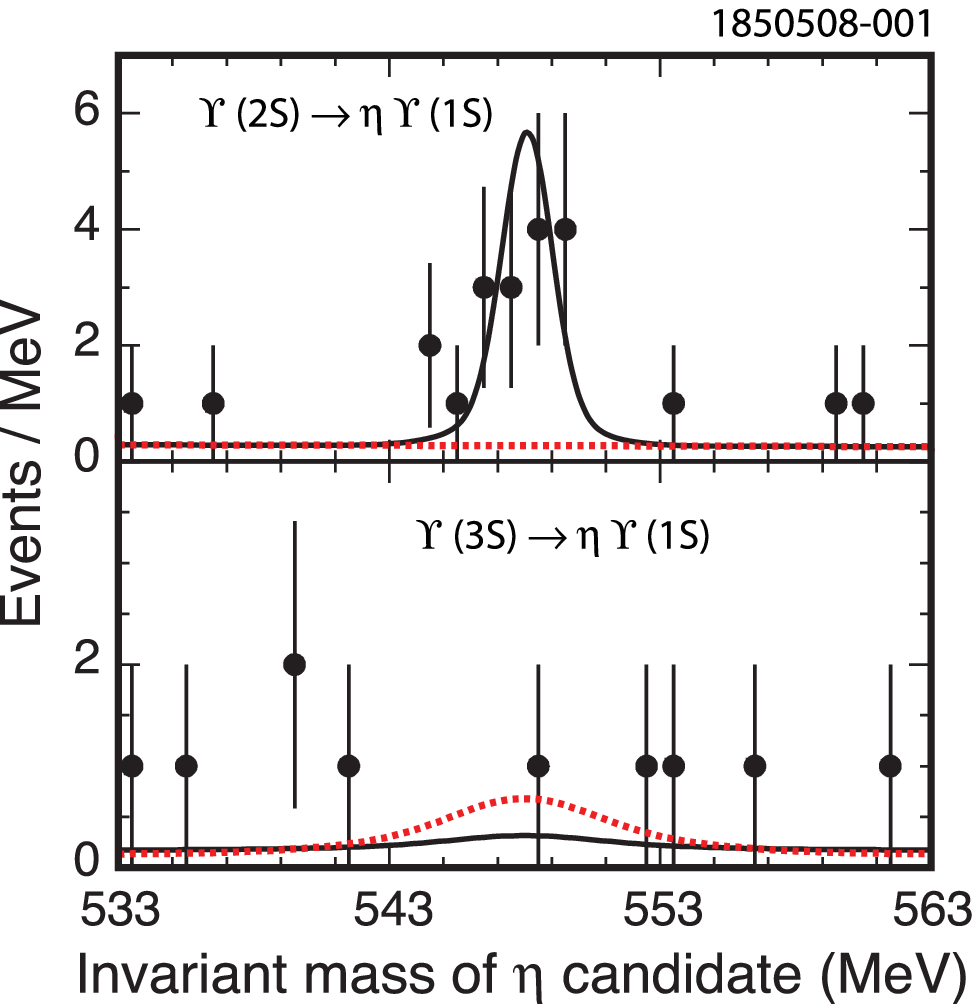}
\includegraphics[width=0.45\linewidth]{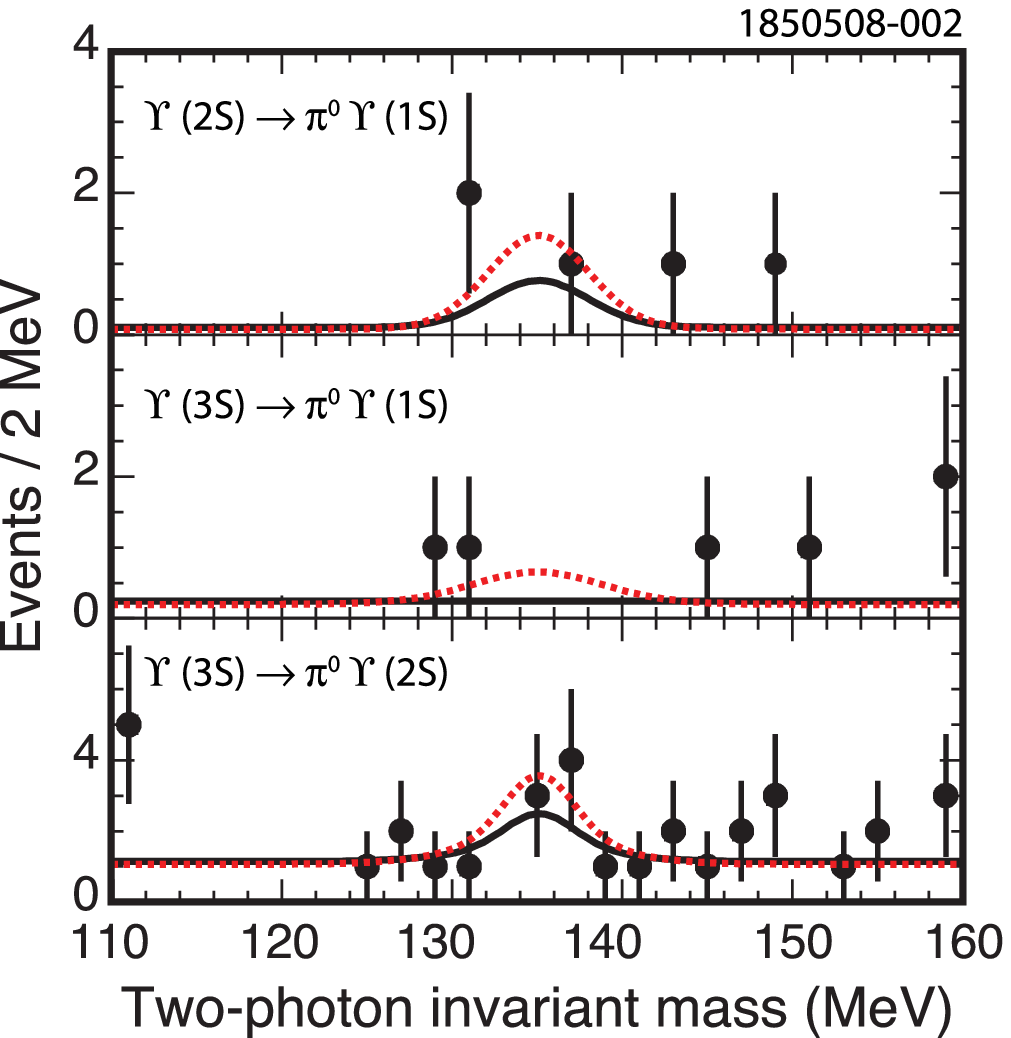}
\caption{Invariant mass of (left) $\eta$ and (right) $\piz$ candidates observed in transitions
$\upsms\goesto\upsns + X(\gamgam,\pipi\piz, 3\piz)$.\label{etatrans}}
\end{wrapfigure}\subsection{$\eta$ Transitions}
We next present the first observation~\cite{etapaper} of a transition in bottomonium involving $\eta$ mesons.
In order to produce a pseudoscalar meson $\eta$ or $\piz$ in
$\upsns\goesto(\eta,\piz)\upsms$ transitions (involving the 
flip of a heavy quark's spin), one quark of the hadronic system
must emit an M1 (magnetic dipole) gluon while the other
emits an M1 or E2 (electric quadrupole) gluon.   
The observation of the spin-flip of a b-quark can shed
light on its chromomagnetic moment.

In this analysis, the daughter $\Upsilon$ state is tagged via its decay to $\ell\plus\ell\minus$.  
Branching fractions are then obtained 
from the invariant mass distributions of the $\piz$ or $\eta$ daughters.  (See Figure~\ref{etatrans})  The backgrounds
from various sources are very small, and expected to be linear in the region
of interest.  We thus obtain: 
\begin{equation}
B(\upsii\goesto\eta\upsi) = (2.1\asy{0.7}{0.6}\pm 0.3)\times 10^{-4}  (5.3\sigma).
\end{equation}  For  the other $\piz$ or $\eta$ transitions 
studied, only upper limits were obtained.

\section{Hadronic Annihilation Decays}

By contrast to the processes discussed above, hadronic 
annihilation of heavy quarkonia is a comparatively high $q^2$ process, and thus they probe
quite different features of QCD.  

\subsection{$\chi_b(1P,2P)$ Decays to Light Hadrons}

CLEO has also, for the first time, observed~\cite{chilighthad} decays of bottomonia into light hadrons.  
Using data taken at the higher vector states $\upsiii$ and $\upsii$, we tag the production of $\chi_b(1P,2P)$ by observation of
the appropriate E1 photons.     We then reconstruct over 650 different exclusive final states, and obtain the
yield from the invariant mass distributions for each.   We observe
fourteen modes in which there are $>5 \sigma$ signals for each of $\chib$ and $\chibp$, having branching fractions
in the range $1-20 \times 10^{-4}$. These results can be of use in validating models
of fragmentation of heavy states, and for exclusive reconstruction of $\eta_b$ and $h_b$. 

\subsection{$\chi_b(1P,2P)$ Inclusive Decays to Open Charm}

We have also 
studied~\cite{chicharm}  inclusive decays of $\chi_b(1P,2P)$ to open charm. For even-J states, we expect that 
hadronic decays occur via $gg$, whereas for the J=1 states, $gg$ is forbidden, and the most probable 
intermediate state is $g+q\bar{q}$.   We may test these expectations 
by seeking decays involving open charm, which would tend to be suppressed for $gg$ 
and enhanced, with an expectation of $\approx 25\%$ of the hadronic rate for the $g + q\bar{q}$ intermediate state.  

In this analysis, for events containing at least one $D^{\zero}$  the spectrum of detected photons 
is fitted to obtain rates of $D^{\zero}$ production from each $\chi_b$ state. 
The ratio ${\cal{R}}$ of the $D^{\zero}$ rate to the total hadronic rate 
(roughly the total width minus the radiative width in each case) is calculated.  For the $J=1$ states, we confirm
theoretical expectations, obtaining:  only were
significant results obtained - and each confirms the expectation of $\approx 25\%$ for the $J=1$ states: 
\begin{eqnarray}
{\cal{R}}(\chi_{b1}(1P)) = (24.8\pm 3.8\pm 2.2\pm 3.6)\%\\
{\cal{R}}(\chi_{b1}(2P)) = (25.3\pm 4.3\pm 2.5\pm 2.4)\%.
\end{eqnarray} These results represent the first measurements for the
$J=1$ branching fractions and offer the opportunity for the refinement of models of $\bbbar$ hadronic annihilation 
decays. 
\section{Radiative Transitions and Decays}

The study of radiative transitions and decays offers a third probe of QCD. 

\subsection{Annihilation of $\jpsi$ to $3\gamma$} 
An important test of QED has been the study of the $3\gamma$ decay of Ortho-positronium, and similarly
the $3\gamma$ decay of ortho-charmonium, $\jpsi$, can serve as a laboratory for
the investigation of the QCD by comparing the rate for this decay to the rates for 
$\gamma gg$, $ggg$ or $\ell^+\ell^-$.  Prior to our observation~\cite{threegampaper}  of this decay 
only Ortho-positronium was known to decay to $\gamgam\gamma$.  

Production of $\jpsi$ was tagged via the process $\psip\goesto\pipi\jpsi$, and events containing three additional
showers in the electromagnetic calorimeter were selected.  Events for which the invariant mass of any pair of these showers
corresponded to $\piz, \eta, \eta^{\prime}$ or $\eta_c$ were removed.   An excess of 24.2 events
is observed on top of expected backrounds.    We thus obtain   
\begin{equation}
B(\jpsi\goesto\gamma\gamgam) = (1.2 \pm 0.3 \pm 0.2 ) \times 10^{-5} (6\sigma),
\end{equation}
with which zeroth order predictions~\cite{kqrr} generally agree, but first-order perturbative QCD corrections are huge;  this
measurement presents a significant challenge, therefore, for theory.

\subsection{Decays of Vector Charmonium to $\gamma$ + Pseudoscalar Mesons}

Naively, one expects that the ratio of decay rates of heavy quarkonia via
$\gamma gg$ to that via $ggg$ to scale as $\alpha/\alpha_S$.  However, this is not
borne out in the charmonium system; a recent CLEO-c measurement 
revealed the ratio for $\psip$ is only half of that for $\jpsi$.  
We have probed this result by searching for decays
of $\psip$ and $\jpsi$ to $\gamma$ + $(\piz,\eta,\eta^{\prime})$ which proceed via $\gamma gg$.    Of particular
interest is the ratio $R_n \equiv B(\psi(nS)\goesto\eta)/B(\psi(ns)\goesto\eta^{\prime})$,
which is expected to satisfy $R_1\simeq R_2$.
Previous measurements  revealed $R_1 = 20.2\pm 2.4\%$ and 
$R_2 < 66\%$ at 90\% CL .

We searched for all the above $\gamma +$ pseudoscalar decays of $\jpsi$, $\psip$ and
$\psi(3770)$ and found~\cite{gammappaper} that 
$R_2 << R_1$ at 90\% CL.  We've tightened the result for $R_1$, with $R_1 = 21.1\pm 0.9\%$, and obtain a much lower 
limit of $R_2 < 1.8\%$ at 90\% CL.  Such a small value of $R_2/R_1$ poses a significant
challenge to our understanding of these decays. 

\subsection{Radiative Production of $\eta_c$ from $\psip ,\jpsi$}

The M1 radiative transitions $(\psip,\jpsi)\goesto\gamma\eta_c$ represent fundamental processes
whose rates serve as important benchmarks for theory, but both are very poorly measured.  
In addition, the partial with measurements of $\eta_c$ are dependent upon these poor 
measurements.   CLEO-c has made new and much improved
measurements~\cite{etac} of each of these branching fractions.

In this analysis, we measure the yield from the inclusive photon spectrum
from $\psip\goesto\gamma\eta_c$ and the yields from the exclusive photon spectra (using identical exclusive final states of $\eta_c$) 
from $\psip\goesto\gamma\eta_c$ and from $\psip\goesto\pipi\jpsi; \jpsi\goesto\gamma\eta_c$.  We obtain 
$B(\psip\goesto\gamma\eta_c)$  from the inclusive photon spectrum, and $B(\jpsi)/B(\psip)$ from the exclusive photon spectra.  Multiplying these two numbers yields 
the branching fraction $B(\jpsi\goesto\gamma\eta_c)$.   We thus find:
\begin{eqnarray}
B(\psip\goesto\gamma\eta_c)=(4.32\pm 0.16\pm 0.60)\times 10^{-3}\;\mbox{and}\\
B(\jpsi\goesto\gamma\eta_c) = (1.98\pm 0.09\pm 0.30)\%.
\end{eqnarray} These are each significantly 
larger than, but much more precise than the previous PDG~\cite{pdg} average, 
and will result in a renormalization of nearly all exclusive $\eta_c$ branching fractions.  
Interestingly the $\eta_c$ masses reported by experiments which observe $\eta_c$ in M1 transitions average 5 MeV 
below those reported
by experiments which produce $\eta_c$ through $\gamma\gamma$ fusion or $\bar{p} p$ annihilation.  In our
study, we observed that depending on the lineshape assumed, we can obtain a mass consistent with {\em either}
M1 transition or direct-production results.   We thus note that a very careful study of the M1 lineshape 
is clearly in order if the $\eta_c$ mass is to be extracted from M1 transitions.

\section{Summary}

\section*{Acknowledgments}
The author thanks the Organizing Committee for a fabulous conference in a gorgeous
setting, and in particular J. Tran Thanh Van for his idea so many years ago
which has brought forth over forty years of fruitful discussion and the wonderful
spirit of Moriond.  The author also gratefully acknowledges the support of National Science Foundation
Grant No. PHY-0603831.

\end{document}